# 1-bit Quantized On-chip Hybrid Diffraction Neural Network Enabled by Authentic All-optical Fully-connected Architecture

Yu Shao[a,c,d,†], Haiqi Gao[a,c,d,†], Yipeng Chen[a,d], Yujie Liu[b], Junren Wen[a,c,d], Haidong He[a,d], Yuchuan Shao[a,c,d], Yueguang Zhang[b], Weidong Shen[b,*], and Chenying Yang[a,b,*]

[a] Hangzhou Institute for Advanced Study, University of Chinese Academy of Sciences, Hangzhou, 310024, China

[b] State key laboratory of Modern Optical Instrumentation, Department of Optical Engineering, Zhejiang University, Hangzhou, 310027, China

[c] Shanghai Institute of Optics and Fine Mechanics, Chinese Academy of Sciences, Shanghai, 201800, China

[d] Center of Materials Science and Optoelectronics Engineering, University of Chinese Academy of Sciences, Beijing, 100049, China

[†] These authors contributed equally to this work.
[*] E-mail: adongszju@hotmail.com (Weidong Shen); ycheny@zju.edu.cn (Chenying Yang)

**Abstract**

Optical Diffraction Neural Networks (DNNs), a subset of Optical Neural Networks (ONNs), show promise in mirroring the prowess of electronic networks. This study introduces the Hybrid Diffraction Neural Network (HDNN), a novel architecture that incorporates matrix multiplication into DNNs, synergizing the benefits of conventional ONNs with those of DNNs to surmount the modulation limitations inherent in optical diffraction neural networks. Utilizing a singular phase modulation layer and an amplitude modulation layer, the trained neural network demonstrated remarkable accuracies of 96.39% and 89% in digit recognition tasks in simulation and experiment, respectively. Additionally, we develop the Binning Design (BD) method, which effectively mitigates the constraints imposed by sampling intervals on diffraction units, substantially streamlining experimental procedures. Furthermore, we propose an on-chip HDNN that not only employs a beam-splitting phase modulation layer for enhanced integration level but also significantly relaxes device fabrication requirements, replacing metasurfaces with relief surfaces designed by 1-bit quantization. Besides, we conceptualized an all-optical HDNN-assisted lesion detection network, achieving detection outcomes that were 100% aligned with simulation predictions. This work not only advances the performance of DNNs but also streamlines the path towards industrial optical neural network production.

# Introduction

In the contemporary era, the world witnesses an unprecedented information explosion, primarily propelled by the rapid advancements in Artificial Intelligence (AI) and the resurgence of deep learning technologies. This phenomenon has catalyzed an exponential surge in computational demands, outpacing Moore's Law with a doubling of computing power requirements every 3.4 months[1]. Such a trend exerts considerable strain on current electronic processors, underscoring a pressing imperative to explore novel hardware architectures. These architectures are essential to bridge the widening gap between the stagnating pace of Moore's Law and the escalating computational demands.

Thanks to the advantages of photon propagation over electron transportation, such as faster speed, lower losses, and stronger interference immunity, the advent of optical neural networks presents a promising solution. These networks boast impressive processing speeds, expansive bandwidths, and reduced energy consumption, positioning them as a formidable alternative to conventional electronic neural networks in the AI domain. Predominantly, optical neural networks (ONN) divide into two categories: On-chip networks, exemplified by the Mach-Zehnder Interferometer (MZI) model[2,3], and spatial light networks, typically represented by the 4f system[4]. Both categories have witnessed extensive research endeavors[2–9].

Recent years have seen the emergence of a groundbreaking optical network framework – the all-optical diffractive deep neural network ($D^2NN$)[10], while the diffraction neural network (DNN), a streamlined variant of $D^2NN$, has garnered considerable interest for its superior throughput and enhanced scalability compared to other optical networks. Ongoing researches continue to delve into the potential of this all-optical network, driving innovations and investigations in areas such as functionality[3,5,6,11–24], integration[7,25–33], and parallelism[34–40].

Existing diffractive neural network architectures, which mainly rely on phase modulation and spatial light diffraction, are bounded by the constraints of the Rayleigh-Sommerfeld model[41]. Such limitations restrict their capacity for arbitrary modulation, a hallmark of fully-connected electronic networks, and consequently impede their performance. The DNNs architecture attempts to mitigate these limitations by increasing the quantity of neurons and the depth of the network. However, this approach inadvertently reduces the modulation effectiveness of individual neurons, leading to a bottleneck in modulation capacity.

To surmount these challenges, we have introduced a hybrid diffraction neural network (HDNN), integrating matrix multiplication—a fundamental operation in Optical Neural Networks (ONNs)—with varied channels, synergistically merged with the DNNs framework. This integration allows our network to maintain the scalability and high throughput intrinsic to DNNs, while significantly enhancing its modulation capabilities, thus bridging the gap between the potential and limitations of DNNs.

Due to the Nyquist sampling theorem, most DNNs currently use metasurfaces as modulation units to avoid aliasing errors, allowing neurons to function accurately. This requirement sets the size of modulation units to be ≤ $\lambda/2n$ ($\lambda$ representing working wavelength and $n$ representing refractive index), which increases the complexity of the fabrication process and the manufacture cost, especially in the visible and near-infrared wavelength range. Additionally, metasurface-based modulation units have low tolerance for diffraction distances and alignment errors between each modulation layer, posing a huge challenge to the integration and scalability of multi-layer networks.

To address this challenge, we innovatively propose a Binning Design (BD) method. This approach involves unifying the modulation parameters within a region to form a modulation unit. Thus, the sampling interval and the size of the modulation unit are no longer equal. This method increases the size of modulation units while reducing the impact of alignment errors and distance offsets on the final performance. The use of BD method not only significantly reduces fabrication costs without affecting performance but also enhances the robustness of the devices. In addition, to simplify the process steps during device integration, we also employed 1-bit quantization during training, where the STE (Straight Through Estimator) algorithm was used to simplify the modulation units of each layer into two types. The two methods mentioned above are of great significance for the practical application of HDNN.

Furthermore, to explore the full potential of HDNN, we developed a HDNN-assisted lesion detection network. This network, by integrating HDNN into a part of the fully-connected architecture, not only achieves high performance but also boasts a markedly enhanced processing speed. Through this network, medical images can be directly inputted, producing output results that indicate the presence of lesions, thereby substantially alleviating the diagnostic workload for medical professionals. This breakthrough has profound implications for the expanded use of optical neural networks in various applications.

In this paper, we unveil two innovative network architectures: the Hybrid Diffraction Neural Network (HDNN) and its On-chip counterpart, the On-chip Hybrid Diffraction Neural Network (On-chip HDNN). The great capability of the HDNN was substantiated through spatial light modulation initially, with experimental findings aligning closely with our simulations, thereby validating our proposed model. Building upon this foundation, we advanced the concept to the On-chip HDNN and achieved successful device fabrication, further reinforcing the efficacy of the HDNN framework and the Binning Design (BD) method. Both architectures exhibit performance that more nearly resembles that of fully-connected networks, significantly reducing the disparity prevalent in traditional DNNs architectures. Moreover, we ingeniously incorporated HDNN into the task of lesion detection, achieving high accuracy in the detection process, which provides valuable insights for the practical application of diffractive neural networks. Our method overcomes challenges posed by metasurfaces constraints in optical diffractive elements, with its effectiveness confirmed through rigorous experimental verification. The implementation of our proposed architectures not only enhances performance but also streamlines the fabrication process, catalyzing the advent of functional innovations and bolstering the practical deployment of future optical neural networks.

**Results**

**HDNN**

In order to enhance the modulation capabilities of optical diffractive neural networks, which were found to be relatively weaker compared to traditional electronic fully-connected networks, and to overcome the challenges associated with their implementation in the visible light spectrum, we proposed the Hybrid Diffraction Neural Network (HDNN). The HDNN architecture, as shown in Fig. 1a, combines phase modulation with multi-channel amplitude modulations, with each channel processing the same inputs but producing different outputs. The channel with the highest output value directly indicates the predicted label, enabling the HDNN to excel in classification tasks following training.

The light propagation in the HDNN is interpreted through the mechanism of diffraction processes with matrix multiplication procedure. As depicted in Fig. 1b, input images, such as digits, are converted into three-dimensional datasets. Each digit initially interacts with the phase modulation layer—a critical step for preserving network operability and promoting scalability, where the phase of the incident light undergoes modification via a Hadamard operation. Once the light traverses the phase layer, it diffuses toward the amplitude layer through a process that essentially embodies convolution with a Fresnel transfer function. The definitive features of this function are determined by the incident wavelength and the distance over which the diffraction propagates. Upon arrival at the amplitude layer, each processed image is flattened and amalgamated into a matrix, with the first dimension representing the pixel count per image and the second dimension aggregating the total number of images. Concurrently, the amplitude layers from various channels are also flattened and merged into a single matrix, where they perform modulation operations within a range from 0 to 1. Though the modulation of the amplitude layers is constrained by the inherent transmittance properties of the materials or structures used, it has been demonstrated that this restriction has a negligible impact on the final network performance[42]. The outputs after amplitude modulation, which are shown in Supplementary Fig. S1, are then concentrated onto a sensor through a lens to realize varying prediction outcomes, realizing matrix multiplication operations essential for augmenting network performance. Leveraging the described methodology, we have effectively incorporated matrix multiplication into the diffractive neural network framework. This innovative adaptation could markedly enhance the accuracy of diffraction neural networks in machine learning applications, establishing the HDNN as a powerful asset in the domain.

It has been widely known that, in conventional optical diffraction neural networks, the precise manufacture of sub-wavelength units presents considerable challenges, particularly within the visible light spectrum, thereby substantially elevating production expenses. Moreover, alignment discrepancies across the modulation layers during light propagation contribute to imperfect experimental outcomes. These issues are exacerbated as sampling precision decreases, further complicating the process. Nevertheless, a large sampling spacing would compromise the accuracy of simulations. To overcome these obstacles, we proposed the Binning Design (BD) method, as shown in Fig. 1c. Using the BD method, we applied consistent parameters to each modulation subunit, with the size of these subunits matching the sampling interval. By aggregating these subunits through binning, we formed the final modulation unit. This innovative approach separates the diffraction sampling interval from the modulation unit. Details regarding the design of the sampling interval in the BD method can be found in the supplementary materials, and the relationship between sampling interval and accuracy was shown in Supplementary Fig. S2. This strategy not only eases fabrication and reduces costs but also enhances computational precision by optimizing the diffraction sampling interval. Through simulations, we established a correlation between prediction accuracy and alignment errors across modulation layers, as depicted in Fig. 1d. We kept the total modulation area constant at 4480 μm and the sampling subunits at 8 μm, while varying the number of modulation units. This adjustment altered the dimensions of each modulation unit, ranging from 8 μm × 8 μm to 80 μm × 80 μm. Analysis of modulation layers of the same area, but with modulation units of varying sizes, clearly demonstrated that the larger modulation unit exhibits reduced sensitivity to alignment errors. Therefore, by employing the BD method, we significantly reduced the experimental complexity, from the perspectives of feature dimension, fabrication tolerance and alignment accuracy.

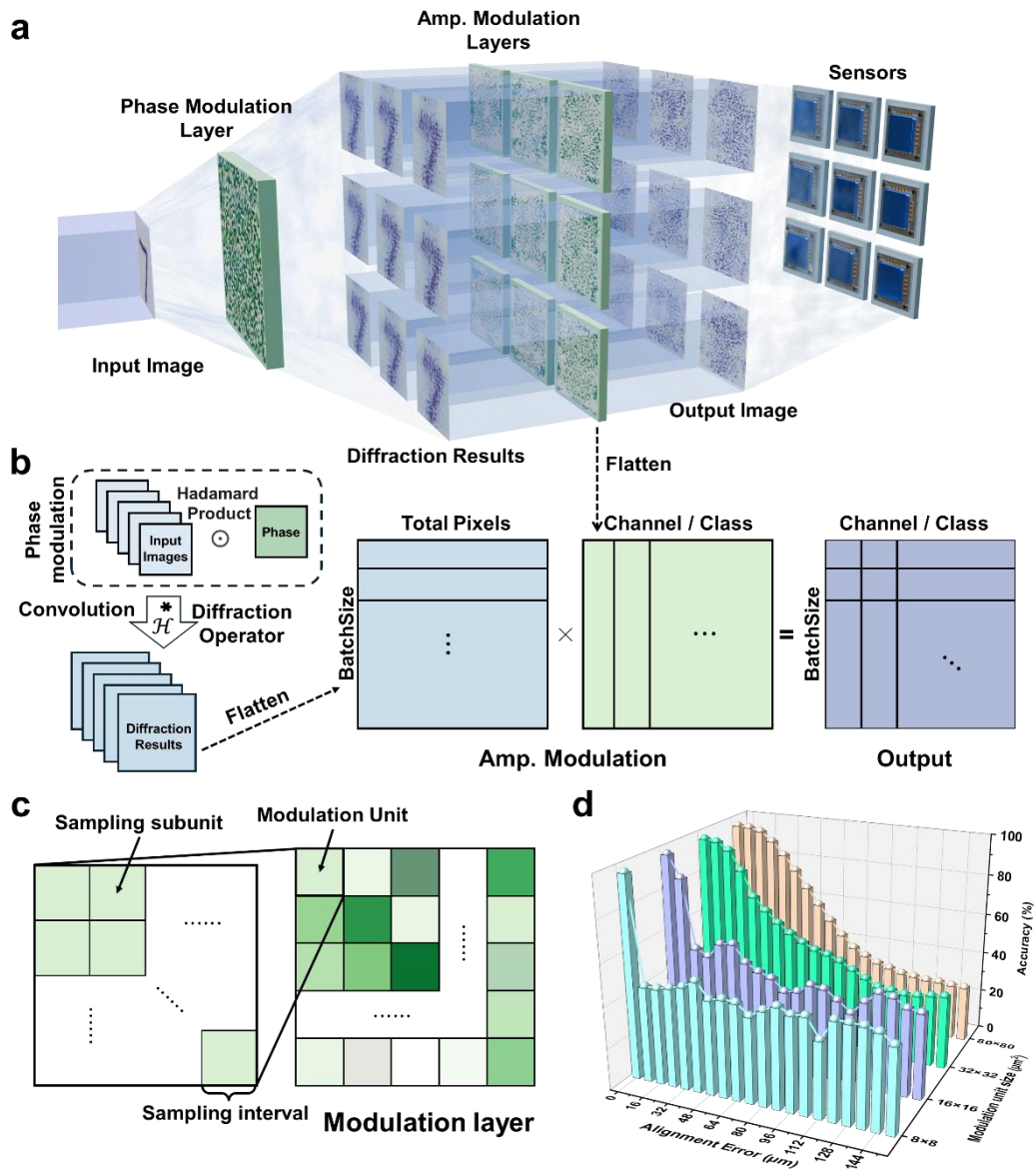

Fig. 1 **Hybrid Diffraction Neural Network and Binning Design method** (a) The structure of the HDNN with one phase modulation layer and one amplitude modulation layer. There are 9 channels in the amplitude modulation layer to produce 9 outputs transporting the ultimate decision-making information. (b) The mathematical perspective of the HDNN. The crucial matrix multiplication is performed during the process of amplitude modulation. (c) The schematic diagram of the Binning Design (BD) method. (d) The relationship between the prediction accuracy and the alignment errors between modulation layers for different sizes of modulation units.

The schematic diagram of the HDNN setup in the experiment is depicted in Figure 2a. At the laser wavelength of 671 nm, we utilized an amplitude spatial light modulator (SLM) with a pixel size of 8 μm × 8 μm to simulate multi-channel amplitude modulation by selecting different display images. Concurrently, we employed a phase-type SLM to carry out the function of phase modulation. The SLMs are all reflective-type and the non-polarizing cube beamsplitters are applied. The detailed setup can be found in Supplementary Material S3. All devices are aligned using a precise method, which was detailed described in Supplementary Material S4, to ensure that the alignment error is within a controllable range. It is noted that while this dual-modulation configuration achieves high prediction accuracy, incorporating additional modulation layers can further enhance performance. With an increase in the number of phase modulation layers, our network evolves into a deep neural network, leading to a slight improvement in the accuracy of digital recognition which effectively demonstrates the scalability of the HDNN, as shown in Fig. 2b. It is pertinent to mention that during this simulation process, to expedite the network training speed, we simplified the network architecture by diminishing the number of neurons in each layer. Despite this simplification, the findings remain insightful. Practically, using the Digital MNIST dataset, we achieved an impressive 96.39% accuracy in simulations (Supplementary Fig. S5) and a remarkable 89% accuracy in real-world experiment with the configuration comprised of the single phase modulation layer combing with the single amplitude modulation layer, shown in Fig. 2a. These results demonstrate the significant advancements we have made in the field, especially the outstanding performance in our experimental setup. Additionally, Fig. 2c vividly demonstrates the correlation between input images and their corresponding output results of both simulations and experiments, showcasing a close alignment of experimental results with simulation predictions. The simulation and experimental confusion matrices are shown in Fig. 2d. Although there are still some deviations from the simulation, this result demonstrates excellent performance beyond traditional DNN devices[43,44]. We hypothesize that the primary discrepancies between simulation and experimentation arise from the non-uniformity and limited collimation of the light source. These factors make it difficult to achieve precise plane-wave illumination, even with the use of a collimating lens. The impact of other parameters in HDNN experiment is also investigated, shown in Supplementary Fig. S6, further demonstrating the excellent error tolerance of HDNN.

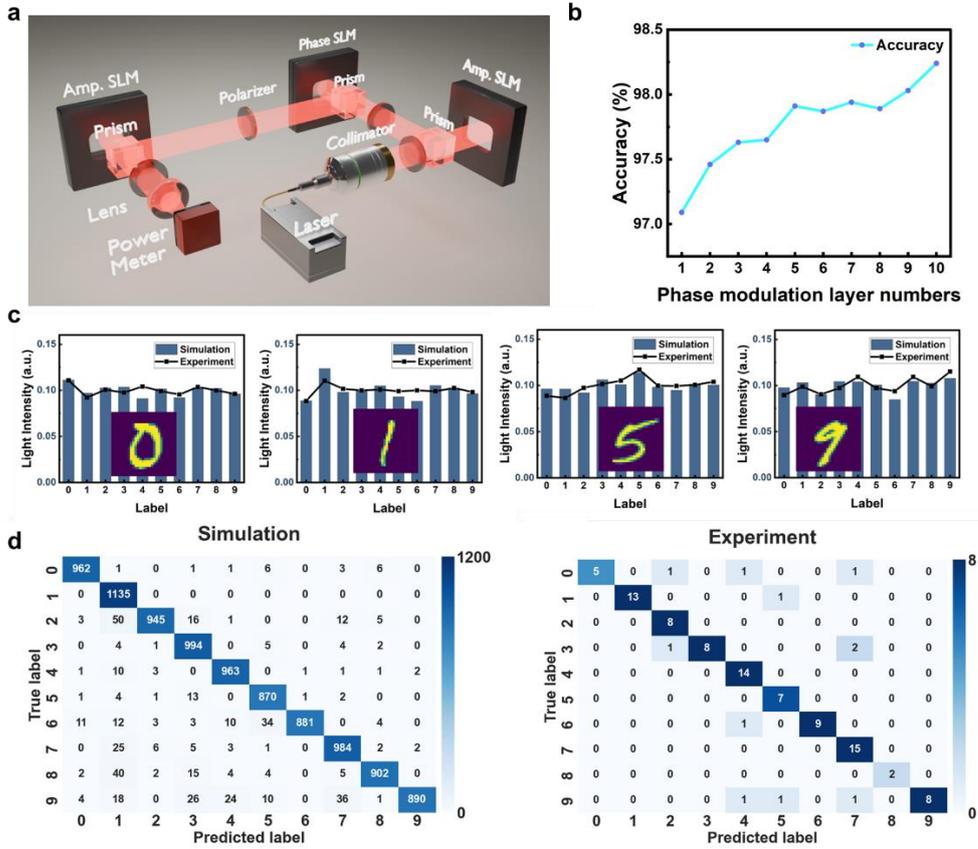

Fig. 2 **Simulations and Experiments of HDNN** (a) The optical path of the HDNN. (b)The influence of the number of phase modulation layers on the final recognition accuracy of the network in simulations. (c) The input image with corresponding output intensity in simulations and experiments. (d) The confuse matrix of the HDNN in simulations and experiments.

**On-chip HDNN**

To highlight the advancements of the HDNN and the BD method, we further proposed the On-chip Hybrid Diffraction Neural Network (On-chip HDNN). Utilizing the BD method enabled us to substitute the metasurfaces for a relief surface, preserving the network's operational integrity. During the training process, we enhanced the error backpropagation process by incorporating 1-bit quantization with the Straight Through Estimator (STE) algorithm, restricting the optimization of modulation parameters to a specific set, as delineated in Fig. 3a. This innovative approach assigns two amplitude modulation values based on the weight $w$: 0 for $w \leq 0$ and 1 for $w > 0$. The adoption of the 1-bit quantization in our network considerably simplifies the fabrication process, while maintaining its functional integrity. The network architecture is exquisitely designed, featuring a precisely arranged grid of $112 \times 112$ modulation units, with each unit covering 5 μm × 5 μm, which was strategically designed for a four-category digit recognition task to demonstrate its capability. To simplify the fabrication process, we confined the phase/amplitude parameter selections for each modulation layer to two distinct values. Consequently, in the actual device fabrication process, each modulation layer requires only two types of relief surfaces. Specifically, the phase modulation layer utilizes phase parameters of 0 and $\pi$, while the amplitude modulation layer employs amplitude parameters of 0 and 100%.

Moreover, a pivotal feature of this model is the additional implementation of beam splitting with the phase modulation layer, a concept inspired by Fresnel Lens and the beam-splitting characteristics of one-dimensional/two-dimensional gratings[45]. The theoretical foundation and computational methods of the design method (Rotary Embedding Method, REM) for the phase beam splitter layer are elaborated in Supplementary Material S7. This unique approach enables the segregation of the input number into multiple regions following their passage through the phase modulation layer and prior to their interaction with the amplitude layer, allowing the On-chip HDNN to achieve snapshot-style digit recognition capability. Importantly, this beam splitting capability extends beyond merely four outputs as we did in the four-category digit recognition task, thereby unlocking the potential for a broader range of functionalities.

As a proof of concept, an On-chip HDNN device was meticulously designed and fabricated with relief surfaces. The detailed preparation steps are outlined in the Supplementary Material S8. Fig. 3b presents the actual photo and the schematic diagram of the On-chip device, illustrating its design and functionality. In the On-chip experiment, the working wavelength is 1064 nm. We employed chromium (Cr) for the digital mask and amplitude modulation layer, and utilized silicon (Si) for the phase modulation layer and the propagation medium. Besides, we added a layer of silicon dioxide film between the phase modulation layer and the digital layer to provide efficient isolation. The etching depth, crucial for precise phase modulation, is calculated using the formula:

$$d = \frac{\lambda}{2(n_{Si} - n_{SiO_2})}$$

where $d$ is the etching depth, $\lambda$ is the working wavelength, while $n_{Si}$ and $n_{SiO_2}$ are the refractive indices of silicon and silicon dioxide at the working wavelength, respectively. The preparation results of the On-chip device are depicted in Fig. 3c, where both Scanning Electron Microscopy (SEM) and Atomic Force Microscopy (AFM) results reveal the precise alignments with our design patterns as well as the excellent verticality of the sidewalls. Additionally, through AFM, we observed that the etch depth of the relief surfaces on the phase modulation layer is very uniform, with a depth of nearly 244nm, which is only 7nm deviation from the designed depth of 251nm, within the tolerance error of preparation. We also investigated the influence of some experimental error factors on the final judgment accuracy, which is presented in detail in Supplementary Material S9, to assist in proving its feasibility. Besides, the patterns illustrating the beam splitting effect at varying exit distances are depicted in Supplementary Fig. S11 to explore the effect of the output distance. Fig. 3d shows a schematic working diagram of the On-chip HDNN device. The device operates with a laser emitting light approximating a plane wave, which passes through a digital mask to form digits. These digits then propagate through the isolation layer to the phase modulation layer. After passing through the forward propagation media, the light is split into several regions whose quantity is equal to the number of channels compacted in the amplitude layer and captured by a common industrial CMOS-based camera. The predicted digit is determined by analyzing the total grayscale value in each region, as shown in Fig. 3d, taking a four-category recognition task as an example. This process is quite straightforward and could be efficiently executed using a simple algorithm.

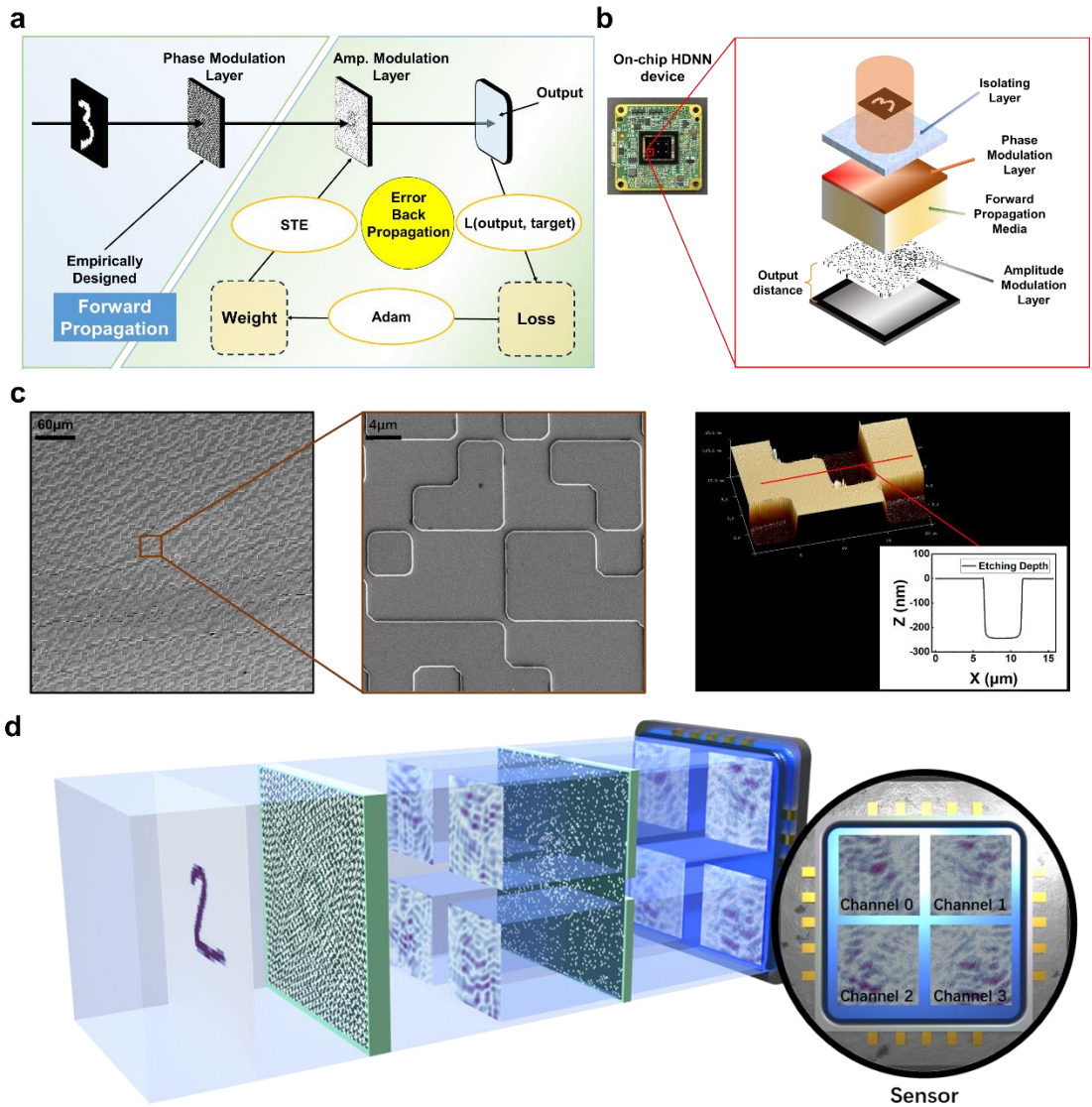

Fig. 3 **On-chip Hybrid Diffraction Neural Network** (a) The schematic diagram of the On-chip HDNN with one phase modulation layer and one amplitude modulation layer. The phase modulation layer was empirically designed with Rotary Embedding Method (REM) to achieve the function of beam splitting. The amplitude modulation layer was trained through 1-bit quantization with the Straight Through Estimator (STE) algorithm. (b) The physical image and structural schematic diagram of the On-chip device. (c) The SEM and AFM images of the critical relief surfaces on phase modulation layer which played a significant role in the integration process of HDNN. (d) Schematic working diagram of the On-chip device with the final result processing. The input image was firstly modulated by the phase modulation layer and split into several regions corresponding to all channels. Then, after the modulation of the amplitude modulation layer, the final result was captured by a CMOS image sensor. The final

result showed in the figure was captured during our experiment, taking four channels as an example.

In the digit recognition task, our On-chip HDNN exhibited an impressive simulation accuracy of 95.43%, reflecting its theoretical effectiveness, while the actual device demonstrated an accuracy of 73.88%. These results are showcased in the confusion matrix, displayed in Fig. 4a, which meticulously presents the differentiation outcomes for various digits. Fig. 4b displays the violin plots of light intensity distributions for each digit, offering a glimpse into the operational intricacies of the On-chip HDNN, revealing the range of intensity variations that the On-chip HDNN processes for accurate digit recognition. Comparative analysis of the simulated and measured light intensity distribution, as shown in Fig. 4c, illustrates the contrast between ideal and experimental performance. Though we can discern the correct digits in experiments, there is a difference in intensity contrast between the intensity of correctly predicted and incorrectly predicted digits, compared to simulations. Furthermore, Fig. 4c also shows both the simulated and experimentally obtained patterns for various digits, where a distinct beam splitting effect is evident. The notable similarity between these two sets of results demonstrates that the device matches well with our initial design objectives, affirming its effectiveness. To enhance the performance of the network further, increasing the number of modulation units could improve the accuracy of number recognition, which was demonstrated in Supplementary Material S11.

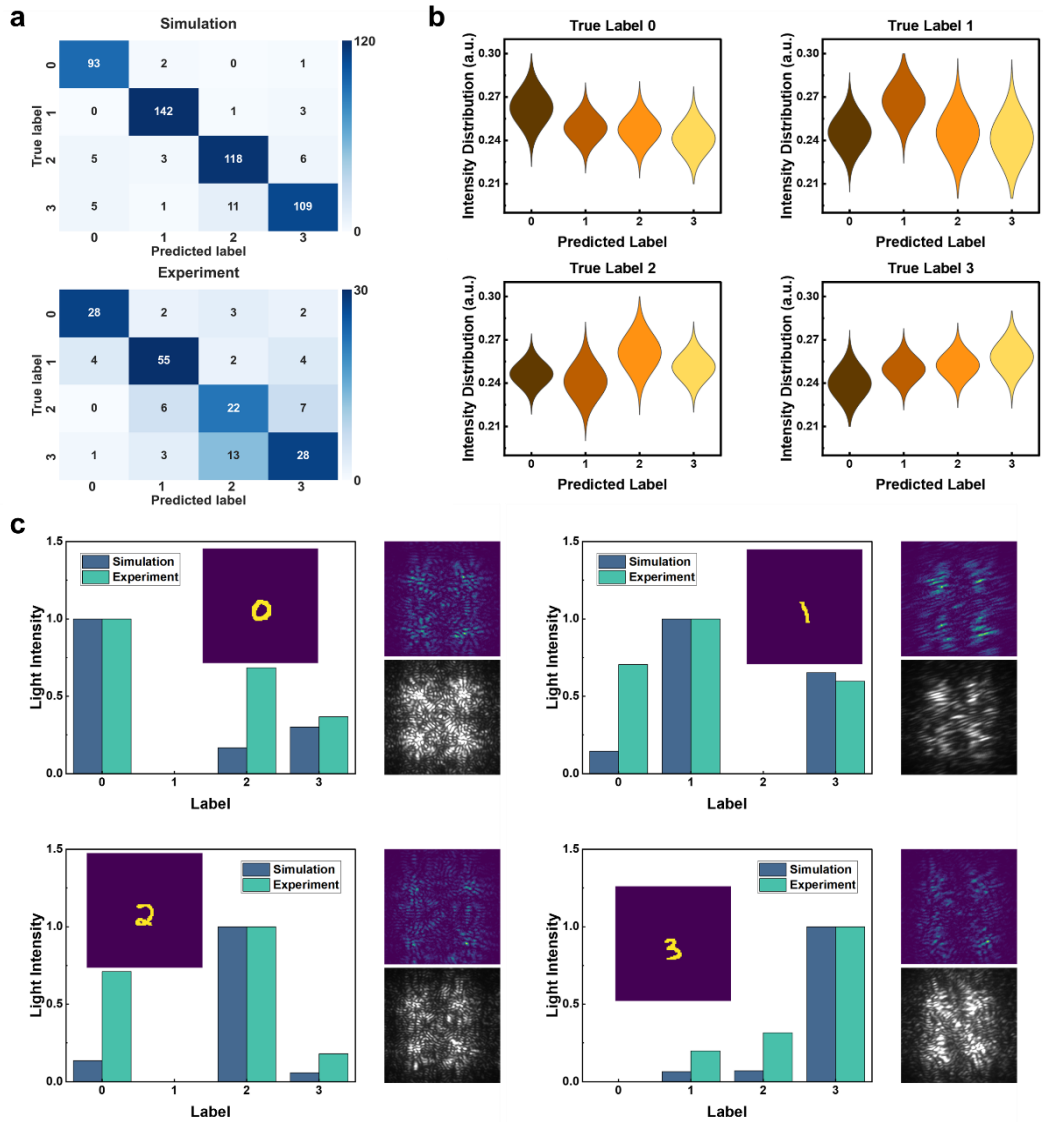

Figure 4 **Simulation and Experiment of On-chip HDNN** (a) The confuse matrices of the On-chip HDNN in simulations and experiments. (b) Experimental intensity distribution maps. (c) The normalized simulated and experimental light intensity distribution for digits 0-3 in different prediction regions, along with their corresponding received simulated (top) and experimental (bottom) images.

**HDNN-assisted lesion detection**

To investigate the capabilities of the Hybrid Diffraction Neural Network (HDNN), we integrated it into a framework for medical lesion detection based on NIH Chest X-ray Dataset[46], provided by National Institutes of Health Clinical Center, as illustrated in Fig. 5a. During the training process, we deliberately selected the same number of images with and without lesions as the training set to ensure the reliability of the training results. The process commences with X-ray images entering the linear neural network, where they firstly undergo a series of convolution operations. The convolutional steps effectively extract features from the images and subsequently resize the outcomes into images with suitable dimensions in the final fully-connected layer. The processed images are then fed into the HDNN as inputs, successively undergoing modulation through the phase layer and the two-channel amplitude layer. Notably, consistent with the foundational principles of optical neural networks, we have deliberately excluded the pooling layers from the convolutional architecture, and the activation function is not used at all in the network. This strategic omission maintains the linear integrity of the entire network, crucial for enabling further all-optical modulation.

Due to practical limitations in materials and devices, we did not fully implement the entire network optically. Instead, we used computer simulations to play the role of the linear neural network, as shown in the Fig. 5b. After obtaining the images processed by the linear neural network, we used the same spatial light device as the digit recognition task for further processing. Compared to the digit recognition task, in the lesion detection task, we utilized two optimizers to optimize the linear network and HDNN separately. Additionally, we introduced two-channel light intensity contrast into the loss function during training. These optimizations facilitated the implementation of our network, culminating in a remarkable 100% agreement between experimental and simulated results for lesion detection. Representative result images from the experimental process are depicted in Fig. 5c. It can be observed that although the predictions are accurate, there are still differences in light intensity contrast between the experimental and simulated results in both channels. We attribute this discrepancy to the low modulation contrast of the SLM, especially in the case of the SLM serving as the amplitude modulation layer, which could be observed in Supplementary Fig. S3b.

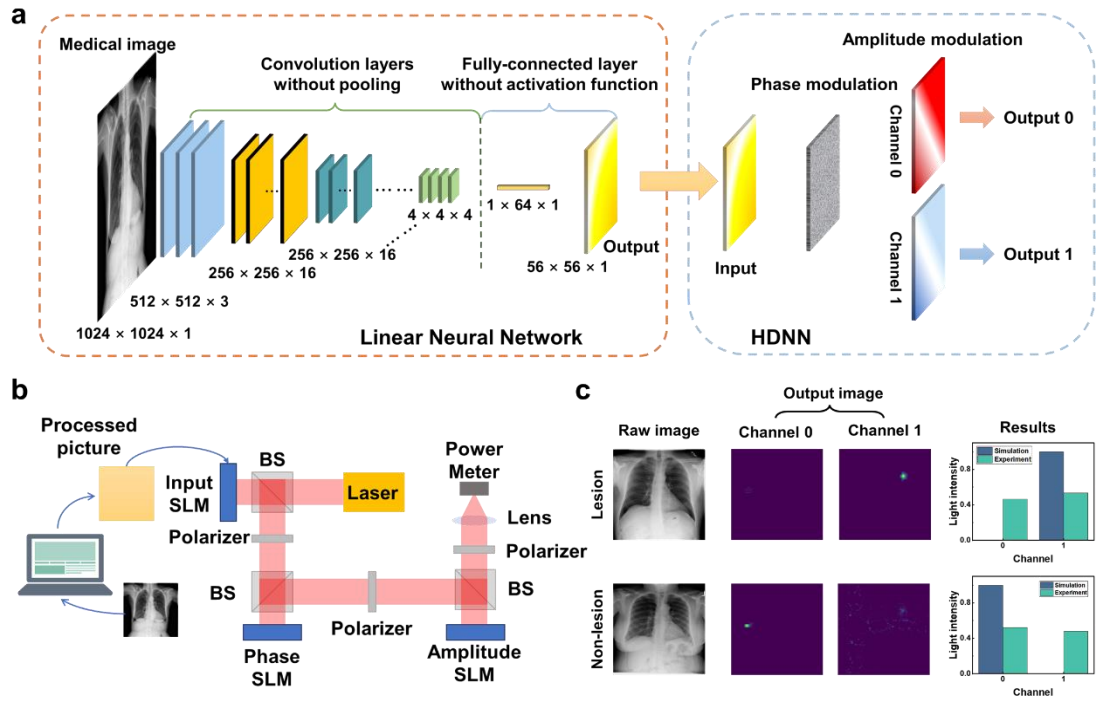

Fig. 5 **HDNN-assisted Lesion Detection** (a) Schematic diagram of network for HDNN-assisted lesion detection. (b) Schematic diagram of the experimental setup. (c) The original X-ray image, its corresponding output image from the linear neural network and the simulated/experimental output light intensity profile of two channels.

**Discussion**

In our work, we initially corroborated the exceptional performance of our newly devised Hybrid Diffraction Neural Network (HDNN), proficient in executing digit recognition tasks via a synergistic combination of a phase modulation layer and an amplitude layer. During this exploration, we discerned that the diffraction process is analogous to a convolution process, with the parameters of the convolution kernel determined exclusively by the diffraction distance. This process eliminates the need for a traditional 4f system in our spatial light modulation setup, thereby streamlining the optical path. Moreover, our spatial light network architecture validated the effectiveness of the Binning Design (BD) method, which opens up possibilities for the wide applications by substantially reducing the preparation/installation costs and difficulties. We established a definitive relationship between sampling interval, wavelength, and structural dimensions. This methodology not only diminishes sensitivity to diffraction distances but also augments tolerance to alignment inaccuracies. Furthermore, owing to the incorporation of matrix multiplication, our proposed HDNN operates with fewer layers and neurons compared to other DNNs, yet it achieves superior accuracy in digit recognition, as described in Supplementary Material S12. Although in our validation process, we only used one layer of phase modulation and one layer of amplitude modulation, the scalability of this architecture goes far beyond that. Naturally, it is easy to build multi-layer deep neural networks similar to $D^2NN$, to achieve a further improved feature. Moreover, during the training process, we did not take into account the contrast of different channels. If the contrast of light intensity between different channels is used as the objective function for optimization, the experimental results would be more visually appealing and would also be significantly less affected by experimental errors.

Advancing from spatial light validation, we integrated the device subsequently and engineered a phase layer with the beam-splitting functionality, thereby integrating all the channels into a single amplitude layer, which is a distinct advantage over the HDNN. Despite the challenge of replicating the exact beam-splitting efficiency of prism-like structures, the digit features were adequately preserved post splitting. Leveraging deep learning, the performance of the network remained robust. We experimentally developed a device capable of four-beam splitting, yet the design can be extended to more intricate beam-splitting phase layers for expanded functionalities. Furthermore, optimizing the beam-splitting layer through deep learning can significantly enhance device performance. In this On-chip HDNN device, we employed two phase modulation values (0 and $\pi$) and two amplitude modulation values (0 and 100%). Apart from simplifying the preparation process, this approach also reduced the tolerance for manufacturing precision, impacting the yield rate of final samples. In industrial production, adopting a variety of phase steps and amplitude steps could substantially elevate the yield rate. Our fabrication process, characterized by lower feature line width requirements and reduced production costs as detailed in Table 1, employs the BD method to substitute metasurfaces with more easily fabricated relief surfaces, making it exceptionally well-suited for mass production. This applies especially to the On-chip HDNN, designed for straightforward chip incorporation, rendering it ideally suited for integration into industrial camera systems.

During the validation experiments of the On-chip HDNN, we also observed that there are some discrepancies of the final light intensity distributions between the fabricated devices and the simulation results, which can be ascribed to the fabrication errors in multiple-step processes. Firstly, through microscopic observation, we found that there is approximately a 1-micrometer offset between the digital layer and the phase layer, as shown in Supplementary Fig. S13, indicating that there will inevitably be misplacement errors in the digital layer and the amplitude layer. The risk of mutual blocking between the two layers of chromium greatly affects the final light intensity output and reduces the contrast. Secondly, limited by the sensitivity and precision of the industrial camera in our imaging experiments, we achieved a grayscale differentiation accuracy of up to 12 bits, whereas in numerical simulations, the bit depth far exceeds 12. This restriction of the image acquisition equipment will lead to the information loss, thereby affecting the prediction accuracy in our experiments. In addition, there is a ~7nm etching error of the silicon depth within the phase modulation layer, which may also affect the performance of our device. It is worth noting that we did not impose any constraints during the training process to intentionally increase the contrast of the predicted light intensity for different numbers. If this is taken into consideration, the tolerance for various fabrication errors during the device fabrication process can be significantly enhanced. Moreover, we can choose slightly larger modulation unit sizes to further reduce alignment difficulties, similar to the design of spatial light validation, which would significantly increase the contrast in the obtained light intensity for different digits.

Besides, to further demonstrate the superior performance of our proposed architecture, we introduced a HDNN-assisted lesion detection network. Due to the principle of linearity in network design, the entire process is amenable to implementation via an all-optical network. Specifically, convolutional tasks can be facilitated by the establishment of 4f optical systems[4], and the integration of the fully connected layer is achievable through the deployment of Diffraction Neural Networks (DNNs). Although undertaking such an all-optical project is ambitious, it promises to fully exploit the inherent advantages of all-optical networks, notably their swift processing capabilities, extensive throughput, and efficient energy consumption.

While our proposed HDNN achieves performance close to that of the fully-connected layer in electronic neural networks, optical diffraction neural networks still face some unresolved challenges. Within our current network architecture, the implementation of nonlinear modulation remains unrealized, primarily due to material selection constraints. This limitation in nonlinear modulation curtails the expansion of our HDNN into broader application domains. As a result, it affects both the modulation efficacy and training speed, preventing these aspects from achieving the same level of excellence found in electronic networks. Addressing this challenge constitutes one of our future directions for innovation. Furthermore, the reconfigurability of the network remains a challenging issue. Once the parameters of an optical diffraction neural network are trained and specified, they cannot be easily altered or replaced like the ANN-driven intelligent metasurface[47], which is more practical in a wider range of scenarios. It is anticipated that optical neural networks will surmount these challenges and attain the ultimate objective of photonic computation in the near future.

**Table 1 Comparison of different networks**

| Work | Network | Wavelength | Modulation Unit | Period | Accuracy | Patterning Method | System Size |
|---|---|---|---|---|---|---|---|
| **Ref. 10** | $D^2NN$ | 750 μm | Metasurfaces | 400 μm | 91.75% | 3D-printing | ~8cm × 8cm × 16cm |
| **Ref. 7** | On-chip DONN | 1.55 μm | SSSD | 500 nm | 90% | - | ~280μm × 250μm × 1010μm |
| **Ref. 28** | MDNN | 532 nm | Metasurfaces | 400 nm | 90.50% | E-beam Lithography | ~112μm × 112μm × 84.2μm |
| **Our work** | On-chip HDNN | 1.064 μm | Relief surfaces | 5 μm | 95.43% | UV Lithography | ~560μm × 560μm × 1400μm |

## Methods

### Simulation

In this study, Fresnel diffraction theorem served as the computational foundation for modeling the diffraction process within the HDNN and On-chip HDNN. For parameter optimization, we leveraged PyTorch (v1.13.0+cu116), using a workstation (GeForce GTX 3090 Ti Graphical Processing Unit, GPU, and Intel (R) Xeon (R) Gold 6248R CPU × 2 @3.00 GHz and 256 GB of RAM, running a Windows 11 operating system). To accurately simulate the physical constraints of the optical components used in the HDNN, we imposed specific limitations on the modulation parameters: phase modulation parameters were constrained within a range of 0 to $2\pi$, and amplitude modulation parameters were limited to a range between 0 and 1. Meanwhile, for the On-chip HDNN, we adopted a binary approach to the modulation parameters: phase modulation was restricted to either 0 or $\pi$, and amplitude modulation was similarly confined to binary states of 0 or 100%. This distinction in parameter constraints reflects the fundamental differences between the two network designs and their respective operational paradigms.

During the optimization chapter, the Adam optimizer was employed due to its efficiency in converging on optimal parameters. The training process utilized a cross-entropy loss function, a standard choice for classification tasks due to its effectiveness in penalizing incorrect classifications. The learning rates were finely tuned to 0.005 for the HDNN and 0.0001 for the On-chip HDNN.

### Optical system of HDNN

In the construction of the HDNN system, critical dimensions and operational parameters were meticulously defined to ensure the integrity and functionality of the proposed structure. The configuration maintained a specific separation of 434 mm between the phase and amplitude modulation components. Spatial Light Modulators (SLMs) utilized across the setup were standardized with a pixel resolution of 8 μm × 8 μm, facilitating fine-grained control over the modulation process. The modulation area for each SLM was set to 8512 μm × 8512 μm, providing ample space for complex optical interactions. The HDNN architecture was designed with optical modulation units sized at 152 μm, organized in a dense matrix consisting of 56 × 56 units. This arrangement allowed for a comprehensive network of optical neurons, capable of performing intricate computational tasks through diffraction-based processes. The diffraction calculation, critical for accurate simulation and prediction, adhered to a sampling interval of 8 μm, matching the pixel size of the SLMs to ensure consistency in the diffraction pattern analysis. The working wavelength is 671nm in the HDNN experiment.

**Configuration of On-chip HDNN**

In constructing the On-chip HDNN, we designed optical modulation units with a size of 5 μm × 5 μm, set within a modulation area of 560 μm × 560 μm. This configuration yields a network of 112×112 optical modulation units. The diffraction calculation employs a 500 nm sampling interval, and the phase and amplitude modulations are separated by 500μm. A high refractive index silicon medium is utilized between the modulations to enhance diffraction effects by increasing the optical path length to 1.77 mm. In order to optimize the diffraction effects during propagation through the Si, we chose a working wavelength of 1064nm for the On-chip HDNN.

**The training process of the network for HDNN-assisted lesion detection**

To enhance our training methodology, we specifically selected 9547 images each of 'no finding' and 'Infiltration' classes from the NIH Chest X-ray Dataset as our training dataset. And the configuration parameters for the HDNN module within this detection network—the size of modulation subunits, the dimensions of modulation units, and their respective quantities—are consistent with those applied in the digit recognition task, with differences emerging solely in the approach to training. Firstly, we engage two separate optimizers (Adam) with different learning rates for the linear neural network and the HDNN. During the training process, an ingenious strategy is implemented. The training process starts with the application of the Cross-Entropy loss function to the network to improve lesion recognition accuracy. Upon the lesion recognition accuracy reaching a threshold of 60%, an alternative loss function focusing on light intensity contrast is introduced. Thus, the cumulative loss, integrating both accuracy and contrast losses, is utilized to train the network comprehensively in the next stage. This methodological approach has culminated in the successful development of a lesion recognition network characterized by both high accuracy and contrast, thereby demonstrating exceptional performance.

**Data availability**

The MNIST dataset is available at http://yann.lecun.com/exdb/mnist/, whereas the NIH Chest X-ray Dataset Source data is available at https://nihcc.app.box.com/v/ChestXray-NIHCC. All other relevant data supporting the findings of this study are available from the corresponding authors on request.

**Acknowledgements**

The authors acknowledge Prof. Ning Zhou and Prof. Xuezong Yang for assisting the experiment. We thank the ZJU Micro-Nano Fabrication Center for the facility support. We extend our gratitude to the National Institutes of Health Clinical Center for the NIH Chest X-ray Dataset and to the creators of the MNIST dataset for their contributions to our research.

**Author contributions**

The manuscript was written through contributions of all authors. CY and WS conceived the study and supervised the project. HG, YS developed the hybrid network and performed simulation, YS, HG, YL, JW fabricated the on-chip device, YS, HG, YC, HH performed the experiments and analyzed the data. HG, YS wrote the paper, which was then discussed with CY, WS, YS and YZ. All authors approved the final version of the manuscript.

**Funding**

This work is the supported by Research Funds of Hangzhou Institute for Advanced Study, UCAS (No. 2023HIAS-Y008).

**Competing interests**

The authors declare that they have no competing interests.

**Additional information**

**Supplementary information** The online version contains supplementary material available at https://doi.org/XXXX.